\documentclass[12pt]{article}

\usepackage{epsfig}
\usepackage{latexsym}

\begin{document}

\begin{titlepage}

\baselineskip 24pt

\begin{center}

{\Large {\bf Fermion Mixing and Mass Hierarchy as Consequences of Mass
Matrix Rotation}}

\vspace{.5cm}

\baselineskip 14pt

{\large Jos\'e BORDES}\\
jose.m.bordes\,@\,uv.es\\
{\it Departament Fisica Teorica, Universitat de Valencia,\\
  calle Dr. Moliner 50, E-46100 Burjassot (Valencia), Spain}\\
\vspace{.2cm}
{\large CHAN Hong-Mo}\\
chanhm\,@\,v2.rl.ac.uk \\
{\it Rutherford Appleton Laboratory,\\
  Chilton, Didcot, Oxon, OX11 0QX, United Kingdom}\\
\vspace{.2cm}
{\large TSOU Sheung Tsun}\\
tsou\,@\,maths.ox.ac.uk\\
{\it Mathematical Institute, University of Oxford,\\
  24-29 St. Giles', Oxford, OX1 3LB, United Kingdom}

\end{center}

\vspace{.3cm}

\begin{abstract}

It is shown that a fermion mass matrix changing in orientation (rotating)
with changing scales can give a simple yet near-quantitative explanation
for quark mixing, neutrino oscillations and the fermion mass hierarchy.

\end{abstract}

\end{titlepage}

\clearpage

\baselineskip 14pt

Like other quantities in quantum field theory such as the familiar running
coupling constant, the fermion mass matrix also varies with changing scales.
That its eigenvalues, namely the mass values, do actually run has already 
been verified experimentally in certain cases \cite{runmass}.  It should 
thus come as no surprise that the fermion mass matrix may also change its 
orientation in generation space (rotate) as the scale changes.  

Indeed, even in the situation when there are no other forces involved than 
those currently studied in the Standard Model, it is easily seen that the 
fermion mass matrices must rotate with changing scales once given nontrivial 
mixing between the $U$- and $D$-fermion states.  For instance, the $U$ mass 
matrix satisfies a renormalization group equation \cite{rge}:
\begin{equation}
16 \pi^2 \frac{dU}{dt} = -\frac{3}{2}D D^{\dagger} U + ...
\label{rgeq}
\end{equation}
which contains a term on the right which is nondiagonal even when $U$ is
diagonal if the $D$ mass matrix is related to $U$ by a nontrivial mixing
matrix.  Hence, as the scale changes, a $U$ matrix diagonal at one scale
will no longer remain so at another scale, or in other words, it will 
``rotate'' with changing scales, as claimed.  And since nontrivial mixing 
has long been established for quarks, while for leptons it has been strongly 
indicated if not already confirmed by recent experiment \cite{SuperK,Soudan},
we conclude that both the quark and lepton mass matrices must rotate with 
changing scales even in the traditional Standard Model scenario.

Looking further afield, there are good reasons to consider the possibility 
of having further, perhaps more direct, mechanisms driving the mass matrix 
rotation.  The mere fact that different generations of fermions can rotate 
into one another, as the mass matrix rotation implies, already means that they
are not distinct entities as once conceived but just different manifestations 
of the same object, like the different colours of a quark, related presumably 
by some continuous ``horizontal'' symmetry \cite{horizontal}.  And if 
this symmetry is gauged, as all other known continuous symmetries seem to be, 
then it would give rise to new forces which can change the generation index 
and hence contribute to the rotation of the fermion mass matrices.  And these 
contributions would be over and above those driven by nontrivial mixing via 
(\ref{rgeq}).

Indeed, what we wish to point out in this note is a, to us, attractive 
possibility that it is the rotation of the mass matrix (due presumably to
some new as yet unfamiliar forces) which is giving rise to fermion mixing 
rather than the other way round.  As we shall show, this can give us an 
immediate explanation not only for the remarkable fermion mixing pattern 
observed in experiment but also for the otherwise puzzling hierarchical 
fermion mass spectrum.

The reason why one can at all entertain this possibility is that once the mass
matrix rotates then it can generate fermion mixing and nonzero masses for the
lower generation fermions even when one starts with neither.  To see this,
one will need first to re-examine some basic premises which have to be
revised in view of the mass matrix rotation.  When the mass matrix has a
scale-independent orientation (i.e. when it does not rotate), the state
vectors in generation space representing the different generations are
trivially defined as the eigenvectors of the mass matrix.  However, if
the mass matrix rotates, then so will its eigenvectors, and the above
definition of the generation states becomes imprecise, for it will need 
to be specified at what scale(s) these states are to be taken as the mass 
eigenvectors.  Consider first as example the $t$ quark, which we can 
take as the eigenstate of the $U$-type quark mass matrix $m^U$ with the 
largest eigenvalue, say $m^U_1$.  But this value $m^U_1$ depends on 
the scale $\mu$, and one usually defines the $t$ quark mass $m_t$ as the 
value at the scale equal to the value itself, i.e. when $\mu = m^U_1(\mu)$.  
It seems natural then to define also the $t$ quark state vector as the 
corresponding eigenvector of $m^U$ also at the same scale.  Similarly, one 
would define the state vector of the $b$ quark as the eigenvector of the 
$D$-type quark mass matrix with the largest eigenvalue taken at the scale 
$\mu = m^D_1(\mu)$.  If so, the state vectors for the $t$ and $b$ quarks would 
be defined at different scales.  Hence, even with the additional ansatz that 
the $U$ and $D$ mass matrices have always exactly the same orientation (i.e. 
aligned eigenvectors) at the same scale, the state vectors of $t$ and $b$ will
be different since the eigenvector would have rotated from the scale at
which the $t$ state vector is defined to the scale at which the $b$ state
vector is defined.  In other words, there will be nontrivial mixing between
the $t$ and $b$ states since the mixing (CKM) matrix element defined as
$V_{tb} = \langle t|b \rangle$ will differ from unity.\footnote{We have
defined the CKM matrix here as the overlap matrix between the triads of
physical state vectors of respectively the $U$ and $D$ fermions, i.e. the
same as for non-rotating mass matrices.  Apart from being conceptually
simple, this definition also corresponds to what is usually actually 
measured in experiment.  For example, the element $V_{tb}$ is inferred
experimentally in $t$ decay from its branching ratio to $b$, where $b$ in 
the final state is identified also by the branching ratios in its decay, 
which means the state vector of the decaying particle is in each case  
evaluated at the scale equal to its mass, as it is done here.  In the 
literature, the CKM matrix is sometimes defined instead as the overlap 
matrix between the eigenvectors of the $U$ and $D$ mass matrices taken 
all at the same scale, which matrix is then scale-dependent, since the 
eigenvectors are themselves scale-dependent.  This alternative definition
is of course perfectly justified if used consistently and is unambiguously 
related to the experimentally measured mixing matrix if it is known how 
the mass matrices rotate.  If the rotation is slow in the energy range
of fermion masses, then the two definitions are nearly equal, but if the 
rotation is of the order considered here, the difference is important,
in which case the definition adopted in this paper seems to us more 
convenient.}

\begin{figure}
\centerline{\psfig{figure=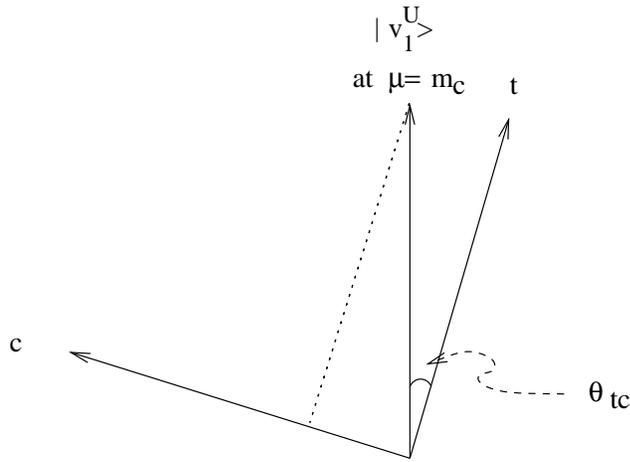,scale=0.95}}
\caption{Obtaining lower generation masses by the ``leakage'' mechanism}
\label{rotationm}
\end{figure}

Similarly, one sees that when the mass matrix rotates, lower fermion 
generations will acquire masses even when they start with none.  To illustrate
the point, it is sufficient to consider only two generations, i.e., say, for 
the $U$ quark case, only the $t$ and $c$ quarks.  We have already defined the 
$t$ state vector above at the scale of the $t$ mass, and since there are only 
2 states, it follows that the $c$ state vector is also uniquely defined as 
the vector orthogonal to $|t \rangle$.  This is necessary since $c$
is quantum mechanically an independent state from $t$.  Suppose now that
the $U$ quark mass matrix at any scale has only one nonzero eigenvalue,
which at the scale $\mu = m_t$ we take to be $m_t$.  It follows therefore
by assumption that the $c$ state taken at the same scale will have zero
mass.  But this should not be taken as the mass of the $c$ quark, for by
the definition above, the physical mass of any fermion state is to be
defined at the scale of the mass itself, namely for the $c$ state it 
should be taken at the scale $\mu = m_c$, not at the scale $\mu = m_t$.
At $\mu = m_c$, however, the sole eigenvector $|v^U_1 \rangle$ of the 
mass matrix $m^U$ with nonzero eigenvalue $m^U_1$ would have already 
rotated to a different direction, as depicted in Figure \ref{rotationm}, 
and acquired a component in the $|c \rangle$ direction giving thus a 
nonzero mass value $\langle c|m^U|c \rangle$ taken at this scale, as
anticipated.  A similar mechanism for generating lower generations masses 
would hold of course for other fermion types, which we shall henceforth 
refer to as the ``leakage'' mechanism.

\begin{figure}
\centerline{\psfig{figure=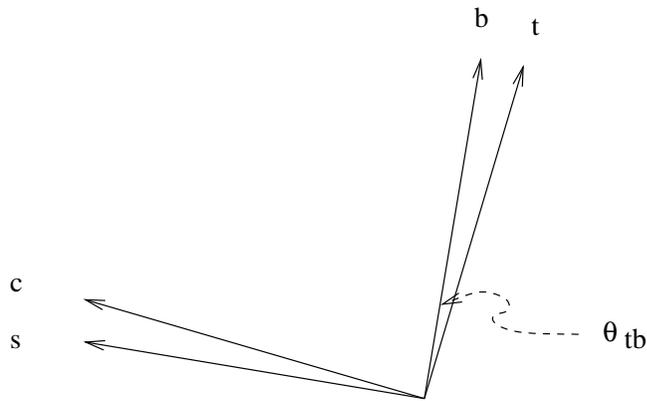,scale=0.95}}
\caption{Obtaining mixing matrices from mass matrix rotation}
\label{rotationV}
\end{figure}

Given that both nontrivial mixing and nonzero lower generation masses
can result from a rotating mass matrix, and both of these effects are
generally small and are in any case otherwise unexplained, it makes sense
to enquire whether they can in fact all be obtained in this way.  One can 
approach the problem empirically since, as we shall show, many of the 
relevant quantities have already been measured and need only to be 
arranged and interpreted in a manner appropriate for the present purpose.
This will be done first in a simplified situation with only 2 generations, 
namely the 2 heaviest, in each fermion-type, which simplification will be 
shown later to approximate already very well the actual 3-generation 
situation.  This makes the analysis much more transparent since the problem 
then becomes planar and there is only one rotation angle and no phases to 
consider \cite{Cabibbo,Jarlskog}.  We have then the pictures shown in Figures 
\ref{rotationV} and \ref{rotationm} for obtaining respectively mixing 
matrix elements and lower generation masses.

Consider first mixing matrix elements.  Suppose from the scale of the $t$
mass to that of the $b$ mass, the mass matrix has rotated by an angle
$\theta_{tb}$.  We have then the dyads of state vectors shown in Figure
\ref{rotationV} for respectively the $U$- and $D$-type quarks.  One 
easily obtains then the CKM elements as: $V_{tb} = \cos \theta_{tb}$ and
$|V_{ts}| = |V_{cb}| = \sin \theta_{tb}$.  From the measured values of these
elements given in the latest databook \cite{databook}, namely:
\begin{equation}
|V_{tb}| = 0.9990 - 0.9993, \ \ |V_{ts}| = 0.035 - 0.043, \ \ 
   |V_{cb}| = 0.037 - 0.043,
\label{VUD}
\end{equation}
one gets thus from each an estimate of the rotation angle, respectively:
\begin{equation}
\theta_{tb} = 0.0374 - 0.0447,\  0.0350 - 0.0430,\  0.0370- 0.0430,
\label{thetatb}
\end{equation}
the values obtained being fully consistent with one another.  (One notes 
that from the same Figure \ref{rotationV}, one could deduce in principle 
also $V_{cs} = \cos \theta_{tb}$, but this will be seen, in contrast to 
the 3 other mixing elements already considered, to be a poor approximation 
receiving large nonplanar corrections when all 3 generations are taken 
into account.)

Consider next the second generation masses obtained by the leakage mechanism.
Suppose from the scale of the $t$ mass to that of the $c$ mass, the mass
matrix has rotated by an angle $\theta_{tc}$, then one sees from Figure 
\ref{rotationm} that $m_c/m_t = \sin^2 \theta_{tc}$.  Hence, from the
measured values of $m_t$ and $m_c$ given in \cite{databook}, namely:
\begin{equation}
m_t = 174.3 \pm 5.1\ {\rm GeV}, \ \ m_c = 1.15 - 1.35\ {\rm GeV},
\label{mtc}
\end{equation}
one obtains the estimate:
\begin{equation}
\theta_{tc} = 0.0801 - 0.0894.
\label{thetatc}
\end{equation}
Similarly, from the measured values from \cite{databook}:
\begin{equation}
m_b = 4.0 - 4.4\ {\rm GeV}, \ \ m_s = 75 - 170\ {\rm MeV},
\label{mbs}
\end{equation}
one obtains the estimate:
\begin{equation}
\theta_{bs} = 0.1309 - 0.2076,
\label{thetabs}
\end{equation}
the error being so large because of the intrinsic uncertainty in defining
the $s$ quark mass, while from the measured values from \cite{databook}:
\begin{equation}
m_\tau = 1.777\ {\rm GeV}, \ \ m_\mu = 105.66\ {\rm MeV},
\label{mtaumu}
\end{equation}
one obtains the estimate:
\begin{equation}
\theta_{\tau \mu} = 0.2463.
\label{thetaumu}
\end{equation}

\begin{figure}
\centering
\hspace*{-2.2cm}
\input{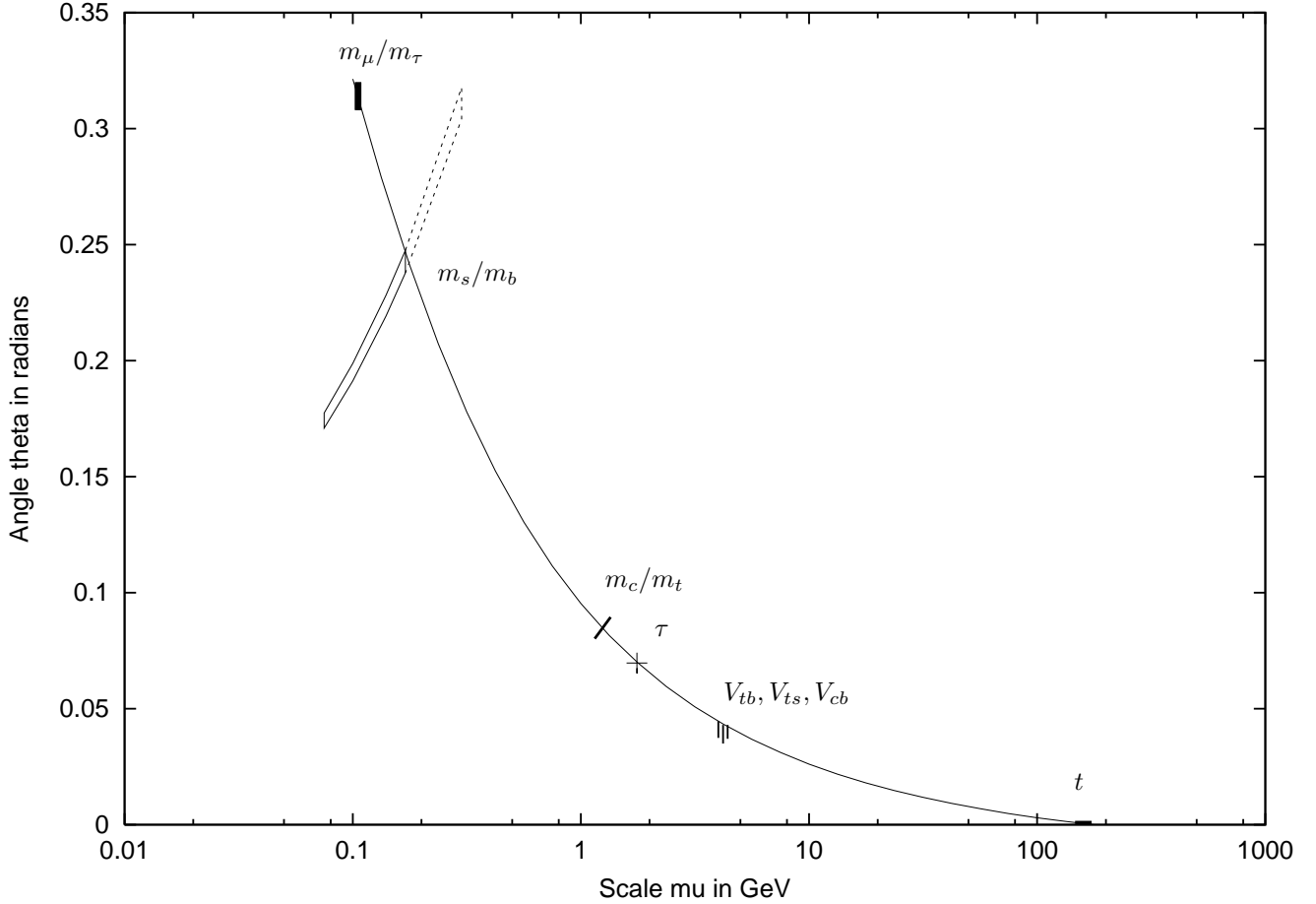}
\caption{The rotation angle in the planar 2-generation approximation, as 
estimated from various mixing elements and mass ratios measured in experiment,
is plotted as a function of the energy scale starting from the $t$ mass.
The error bars shown correspond to the errors in the empirical values of
the mixing elements and masses quoted in \cite{databook}, where in the error 
for the $s$ quark (particulary large because of the intrinsic difficulty in
defining the $s$ mass) the solid bar represents the mass range defined at 
a scale of 2 GeV \cite{databook} while the dashed bar that defined at a
scale of 1 GeV \cite{databook96}.  The curve shown is the result of an 
earlier calculation \cite{phenodsm} with the DSM scheme as detailed in text.}
\label{thetaplot}
\end{figure}

Assume now that the mass matrices of the $U$ and $D$ quarks as well as 
the charged leptons are all aligned at the same scale as proposed above,
and plot the values of the rotation angles obtained before, all starting 
from the direction of the $t$ quark state.  One obtains then the Figure 
\ref{thetaplot} where, in the planar approximation,
we have taken $\theta_{ts} = \theta_{tb} + \theta_{bs}$, and 
$\theta_{t \mu} = \theta_{t \tau} + \theta_{\tau \mu}$, with $\theta_{tb}$
taken from (\ref{thetatb}) and $\theta_{t \tau}$ (indicated by a cross
in Figure \ref{thetaplot}) estimated by interpolation between the values 
of $\theta_{tb}$ and $\theta_{tc}$ given above.   One sees that all the 
gathered information can indeed be comfortably explained by a mass matrix 
rotating smoothly as the scale changes, as suggested.

The above analysis was done under the simplifying assumption of there being
only 2 generations of fermion states, but we shall now show that it is 
already a very good approximation to the actual 3-generation situation.
When 3 generations are introduced, the mixing matrix can be parametrized
as:
\begin{equation}
\left( \begin{array}{ccc} V_{tb} & V_{ts} & V_{td} \\
   V_{cb} & V_{cs} & V_{cd} \\ V_{ub} & V_{us} & V_{ud}
\end{array} \right) =
\left( \begin{array}{ccc} c_1 & -s_1 c_3 & -s_1 s_3 \\
   s_1 c_2 & c_1 c_2 c_3 - s_2 s_3 e^{i \delta} & c_1 c_2 s_3 + s_2 c_3
     e^{i \delta} \\
   s_1 s_2 & c_1 s_2 c_3 + c_2 s_3 e^{i \delta} & c_1 s_2 s_3 - c_2 c_3
     e^{i \delta}
   \end{array} \right).
\label{VKM}
\end{equation}
Notice that although this parametrization, which is more convenient for 
our purpose here, is formally the same as the original Kobayashi-Maskawa 
parametrization \cite{KM}, the rows and columns are labelled differently,
namely in order of decreasing mass rather than in order of increasing mass, 
so that the meanings of the angles are also different.  But, as for the
original Kobayashi-Maskawa parametrization, (\ref{VKM}) can be interpreted 
as a product of 3 Euler rotations and a phase change \cite{Donlowich}:
\begin{equation}
\left( \begin{array}{ccc} 1 & 0 & 0 \\ 0 & c_2 & - s_2 \\ 0 & s_2 & c_2
   \end{array} \right)
\left( \begin{array}{ccc} c_1 & - s_1 & 0 \\ s_1 & c_1 & 0 \\ 0 & 0 & 1 
   \end{array} \right)
\left( \begin{array}{ccc} 1 & 0 & 0 \\ 0 & 1 & 0 \\ 0 & 0 & - e^{i \delta}
   \end{array} \right)
\left( \begin{array}{ccc} 1 & 0 & 0 \\ 0 & c_3 & s_3 \\ 0 & - s_3 & c_3
   \end{array} \right),
\label{Eulerrot}
\end{equation}
where $c_i$ and $s_i$ are the cosines and sines of the Euler-like angles 
$\theta_i$.  One sees then that if we continue to denote as before $V_{tb}$ 
as $\cos \theta_{tb}$, the elements $V_{ts}$ and $V_{cb}$ are no longer 
just given by $\sin \theta_{tb}$ but by respectively $\sin \theta_{tb} 
\cos \theta_3$ and $\sin \theta_{tb} \cos \theta_2$.  However, the 
angles $\theta_2$ and $\theta_3$, being still in the present picture 
just rotation angles undergone by the triad of state vectors from the 
scale $m_t$ to $m_b$, are expected to be small, namely of the order
of the difference in scale times the rotation rate.  Indeed, their actual
values can be estimated from the empirical values given in \cite{databook}
for the corner elements of the CKM matrix in comparison to the values of
$V_{ts}$ and $V_{cb}$ quoted above, giving:
\begin{eqnarray}
|V_{td}| = 0.004 - 0.014 & \longrightarrow & |\tan \theta_3| = 0.093 - 
   0.400, \nonumber \\
|V_{ub}| = 0.002 - 0.005 & \longrightarrow & |\tan \theta_2| = 0.047 - 
   0.135,
\label{tan23}
\end{eqnarray}
from which one gets:
\begin{equation}
\cos \theta_2 = 0.999 - 0.991; \ \ \cos \theta_3 = 0.996 - 0.928.
\label{cos23}
\end{equation}
Hence, one concludes that in the 2 generation planar approximation of
Figures \ref{rotationV} and \ref{thetaplot} where one puts $V_{ts} 
= V_{cb} = \sin \theta_{tb}$, one has made an error of at most a few 
percent.

A similar error has been made in Figure \ref{thetaplot} as regards the
lower generation masses obtained from the ``leakage'' mechanism.  The 
estimate (\ref{thetabs}) is for the angle rotated between the scales of 
$m_b$ and $m_s$ but in Figure \ref{thetaplot} we have added this angle
to the rotation angle from scale $m_t$ to scale $m_b$ to get the angle
from scale $m_t$ to scale $m_s$.  Such an addition is valid in the 2
generation approximation but has nonplanar corrections in the actual
3 generation situation.  In this case, there does not seem to be enough 
empirical information to evaluate the error directly, but its rough value
can be inferred.  The angle between the plane defined by the $t$ and $c$
vectors and the plane defined by $b$ and $s$ is given by the angle between
their normals, namely the vectors for $u$ and $d$ respectively, which
according to \cite{databook} takes the value:
\begin{equation}
|V_{ud}| = \cos \theta_{ud} = 0.9742 - 0.9757 
\label{thetaud1}
\end{equation}
giving
\begin{equation}
\theta_{ud} = 0.2209 - 0.2276.
\label{thetaud2}
\end{equation}
The nonplanar error incurred in the angle at scale $m_s$ plotted in Figure 
\ref{thetaplot} is of order $\cos \theta_{ud}$ and is thus of order 
a few percent, which is negligible given the large error already inherent 
in the definition of the $s$ quark mass.  A similar error is presumably 
present in the angle plotted in Figure \ref{thetaplot} at scale $m_\mu$, 
but one has at present no means for directly ascertaining this.

Thus, as far as the evidence goes in Figure \ref{thetaplot}, one seems
justified in suggesting that both quark mixing and the second generation 
masses of quarks and charged leptons can arise from a continuous rotation 
with scale change of the fermion mass matrix.  The analysis can in principle 
be extended further to lower scales to examine the masses of the $u$ and 
$d$ quarks, and eventually even the electron and neutrino masses and 
lepton mixing, i.e. neutrino oscillations.  For doing so, however, analyses
with the full 3 generation rotation matrices involving 3 angles and a phase
as detailed in (\ref{VKM}) is necessary, and there is as yet insufficient 
empirical information to do so without relying on some extrapolation model.

For this reason, let us consider the curve shown in Figure \ref{thetaplot}.
This appears to be the best-fit to the data points but is actually the
result of a calculation \cite{phenodsm} done two years ago with the Dualized 
Standard Model (DSM) scheme that we ourselves advocate.  The DSM scheme first
suggests a possible explanation for 3 fermion generations as the broken
dual symmetry to colour, and then goes on to attempt a semi-quantitative
understanding of the fermion mass and mixing patterns with reasonable
success \cite{dualgen}.  To achieve the latter purpose, it assumes that
the fermion mass and mixing patterns both arise as consequences of the
rotation of the mass matrix, exactly in the manner suggested by the above
empirical analysis.  The mass matrix rotation in this scheme is driven
by a dual Higgs mechanism the full details of which can be found in e.g.
\cite{phenodsm,dualgen} but need not bother us at this juncture.  We note 
here only the following two points of particular relevance to the present 
discussion.  (a) The mass matrix has a high energy rotational fixed point 
at infinite scale with its heaviest eigenvector pointing in the direction 
$(1, 0 ,0)$ and a low energy fixed point at zero scale with the heaviest 
eigenvector pointing in the direction $\frac{1}{\sqrt{3}}(1, 1, 1)$.  
(b) The rotation of the mass matrix between the two fixed points calculated
to one-loop order depends on 3 parameters, namely 2 Higgs vev's and a 
Yukawa coupling, which were fitted to the data on fermion mass and mixing 
parameters, specifically in \cite{phenodsm} to $m_c/m_t$, $m_\mu/m_\tau$,
and the Cabibbo angle.  

The fact (a) that there is a rotational fixed point at infinite scale means 
that the rotation will go slower at high energy, which is apparently what 
is indicated by the data in Figure \ref{thetaplot}.  This is one reason
why the DSM curve in Figure \ref{thetaplot} is able to reproduce so well 
the scale dependence of the mass matrix rotation with only 2 parameters 
(the third having to do with the degree of nonplanarity as measured by
the Cabibbo angle $\sim \theta_{ud}$).  Conversely, the fact that the 
data points in Figure \ref{thetaplot} by themselves already seem to 
follow a continuous rotation curve also goes some way towards explaining 
the at first sight somewhat puzzling numerical success of the DSM in 
fitting, seemingly so effortlessly, the empirical mass and mixing patterns.

That being the case, it seems worthwhile to explore even lower scales by 
attempting an extrapolation with the DSM formula.  This has already been done 
for the lowest generation masses $u, d$ and $e$ \cite{phenodsm}.  The results 
fall into a sensibly hierarchical pattern but are numerically inaccurate.  
Presumably, this means that the rotation curve calculated to one-loop order 
in the DSM, which has apparently worked over an already surprisingly large 
range of scales down to the $\mu$ mass because of angular proximity to the 
rotational fixed point at infinite scale, is no longer reliable further down.
Nevertheless, it may still be worthwhile to consider with this extrapolation
lepton mixing or neutrino oscillation, where even a qualitative answer would 
be instructive.   Although in so doing an even further extrapolation is needed,
there is, as we shall see, a saving grace by virtue of the low energy fixed 
point (a) at zero scale.  Following then the previous precepts, one should 
presumably define the state vectors of neutrinos at their respective mass 
scales as for the other particles.\footnote{In \cite{nuos,phenodsm}, we had 
defined instead the state vectors of neutrinos at the scales of their 
respective Dirac masses, with which prescription we are thus now at variance. 
As we shall see, however, the result for the MNS elements are qualitatively 
similar, and are equally consistent with existing data, the reason being that
the physical masses of neutrinos as well as their Dirac masses as found in 
\cite{nuos,phenodsm} are all already close to the fixed point at zero scale.  
From the DSM point of view, the present prescription means that the MNS 
matrix elements can now be predicted from the trajectory fitted with quark 
and charged lepton data without any further input from neutrinos; it also 
removes the requirement previously found necessary in \cite{nuos,phenodsm}
of selecting only the vacuum solution to the solar neutrino problem.  
Details of these and other implications on the DSM will, we hope, be 
reported later in a separate communication.}  In particular, for the 
heaviest neutrino $\nu_3$, experiment suggests a mass of around 0.05 eV 
\cite{SuperK,Soudan}, which is some 9 orders of magnitude below the last 
point at the scale $m_\mu$ shown in Figure \ref{thetaplot}.  This means 
first that neutrino mixing angles can be considerably larger than those 
for quarks, which agrees with what has been observed experimentally, but 
it also means that without an accurate extrapolation formula, any estimate 
would seem at first sight quite hopeless.
 
Fortunately, with the DSM scheme, the extrapolation is saved by the fact
that the mass matrix rotation has a fixed point at the scale $\mu = 0$,
which means that the rotation is at least asymptotically bounded.  Further,
at $\mu = m_{\nu_3} \sim 0.05\ {\rm eV}$, one is likely to be already so near 
this low energy fixed point that one can safely approximate the state vector 
of the heaviest neutrino $\nu_3$ just by the vector at the fixed point, 
namely: $\frac{1}{\sqrt{3}} (1, 1, 1)$.  That this is indeed the case is
confirmed by the calculation in e.g. \cite{phenodsm}.  Putting in then the 
state vectors obtained before in \cite{phenodsm} for $\mu$ and $e$:
\begin{eqnarray}
|\mu \rangle & = & (- 0.075925,\  0.774100,\  0.628494), \nonumber \\
|e \rangle   & = & (0.027068,\  - 0.628482,\  0.777354),
\label{muevectors}
\end{eqnarray}
one obtains immediately:
\begin{eqnarray}
U_{\mu3} & = & \langle \mu|\nu_3 \rangle = 0.7660 \nonumber \\
U_{e3}   & = & \langle e|\nu_3 \rangle = 0.1016,
\label{Umue3}
\end{eqnarray}
where one notes that the vector $|e \rangle$, being defined as the vector
normal to both $|\tau \rangle$ and $|\mu \rangle$, is already determined 
at the $\mu$ mass scale, and does not therefore suffer from the inaccuracy
mentioned above in the determination of the $e$ mass through extrapolation.
These numbers, being near maximal for the ``atmospheric'' angle $U_{\mu 3}$
and small for the ``Chooz'' angle $U_{e3}$, are well consistent with the
present experimental limits of around $0.56 - 0.83$ and $0.00 - 0.15$ for
$U_{\mu3}$ \cite{SuperK,Soudan} and $U_{e3}$ \cite{Chooz} respectively.
We note that these predictions are quite robust.  Although the actual 
numbers in (\ref{Umue3}) come from the state vectors (\ref{muevectors})
of $\mu$ and $e$, which depend on the details of the DSM calculation in
\cite{phenodsm}, the conclusion that $U_{\mu 3}$ is large and $U_{e 3}$
small is a consequence only of the ``leakage'' mechanism illustrated by 
Figure \ref{rotationm} which dictates that the $\mu$ state vector should
point in the direction of rotation.  And so long as the rotation from
the $\tau$ scale to the $\mu$ scale is in the general direction of the
asymptotic rotation between the two fixed points at infinite and zero 
scales, the $\mu$ vector will lie roughly on the plane containing the
two asymtoptic vectors and the $e$ vector be roughly normal to this plane, 
from which fact alone the qualitative conclusion that $U_{\mu 3}$ is large 
and $U_{e 3}$ small will already follow.  However, if one goes one step 
further and approximates the state vector of the second heaviest neutrino 
by the tangent to the rotation trajectory at the fixed point, which now  
depends on how the fixed point is approached by the extrapolated rotation 
trajectory and therefore can be inaccurate, one obtains a rough value 
also for the ``solar'' angle $U_{e2}$ of about 0.24, but this lies outside 
the present experimental range of around $0.4 - 0.7$.

Thus it seems that just with the general idea of mass matrix rotation, plus
the two fixed points at infinite and zero scales as suggested by the DSM 
scheme but without the injection of its other details, one obtains already 
a consistent near-quantitative description of both quark and lepton mixing 
plus the fermion mass hierarchy, with most of the salient features (except
for the moment CP-violation) included.  The DSM comes in, apart from 
locating the fixed points, only in supplying a raison d'etre for the 3 
generations of fermions in the first place, and then an explicit theoretical 
mechanism for driving the mass matrix rotation, both of which are of course 
conceptually very important, but are not really essential for deriving the 
above numerical results.

An obvious next question is whether such a description can be independently
tested, and the answer is that perhaps it can, which we consider one of its
attractive features.  Once the mass matrix rotates, then its eigenstates 
defined at one scale will in general no longer be eigenstates at some other 
scale.  In particular, this means that the fermion flavour states defined 
each at its own mass scale as outlined above will not be diagonal states of 
the mass matrix at an arbitrary energy.  Since reaction amplitudes depend
in general on the fermion mass matrix, it follows that these too may become
non-diagonal and lead, as has been suggested in \cite{impromat}, to new 
flavour-violating reactions differing in nature from those arising from, 
say, flavour-changing neutral currents.  Such so-called ``transmutation'' 
effects have been studied in \cite{impromat,photrans} and in other cases
and it is found that with a mass matrix rotating at the sort of speed 
indicated 
in Figure \ref{thetaplot} they could be of sufficient size to be observable 
in some high sensitivity experiments in current operation such as Bepc, 
Cleo, BaBar, and Belle \cite{Bepc,Cleo,Babar,Belle}.  However, although
the observation of flavour-violation at the estimated magnitude favours
its interpretation as coming from a rotating mass matrix as here advocated,
the absence of the same is less conclusive since reaction amplitudes 
depend on quantities other than the fermion mass matrix, such as  
interaction vertices, which may also rotate giving rise to other effects 
modifying the above-cited estimates.  Nevertheless, in view of the very
significant implications on fermion properties discussed in this note, 
an experimental search for such possible transmutation effects due to a 
rotating mass matrix will seem to be an extremely worthwhile quest.

\end{document}